\documentclass[a4paper,12pt]{article}
\usepackage{epsfig}
\usepackage{graphicx}
\usepackage{subfigure}
\begin{document}
\renewcommand{\baselinestretch}{01.35}
\renewcommand{\arraystretch}{0.666666666}
\parskip.2in
\newcommand{\hs}{\hspace{1mm}}
\newcommand{\nhat}{\mbox{\boldmath$\hat n$}}
\newcommand{\cmod}[1]{ \vert #1 \vert ^2 }
\newcommand{\mod}[1]{ \vert #1 \vert }
\newcommand{\pr}{\partial}
\newcommand{\fr}{\frac}
\newcommand{\ie}{{\em ie }}
\newcommand{\p}{\varphi}
\newcommand{\x}{\xi}
\newcommand{\xb}{\bar{\xi}}
\newcommand{\Vd}{V^{\dagger}}
\newcommand{\Ref}[1]{(\ref{#1})}
\title{\hbox{\hspace{18mm}{
\vbox{\Large{\bf Scattering of sine-Gordon Breathers}
\break
\Large{\bf on a Potential Well}}
}}}

\author{
B. Piette\thanks{e-mail address: B.M.A.G.Piette@durham.ac.uk}
and W.J. Zakrzewski\thanks{e-mail address: W.J.Zakrzewski@durham.ac.uk}
\\ Department of Mathematical Sciences,University of Durham, \\
 Durham DH1 3LE, UK\\
}\date{May 2006}

\maketitle

\begin{abstract}
We analyse the scattering of sine-Gordon breathers on a square potential well.
We show that the scattering process depends not only on the vibration 
frequency of 
the breather and its incoming speed but also on its phase as well as the 
depth and width of the well.
We show that the breather can pass through the well and exit with a 
speed different, sometime larger, from the initial one. It can also be trapped 
and very slowly decay inside the well or bounce out of the well and go back 
to where it came from. We also show that the breather can split into a kink 
and an anti-kink pair when it hits the well.

\end{abstract}

\section{Introduction}
The sine-Gordon model is probably the most studied integrable model.
One of the reasons for this is that it describes a large variety of
physical systems ranging from the Josephson effects \cite{Josephson},
particle physics \cite{Skyrme}, information transport in
microtubules \cite{Abdalla}, non-linear optics \cite{McCall},
and ferromagnets \cite{WBK}. 

From a mathematical point of view it is interesting for several reasons.
First of all it is Lorentz invariant. This means that any stationary solution 
can be boosted to any speed, a key property to perform any scattering.
Moreover as any finite energy solution, say, describing a kink, 
of the sine-Gordon model corresponds to 
a mapping from the circle into itself, each solution 
is characterised by a topological charge taking integer values. 
By conventions, solitonic solutions
with a negative topological charge are called anti-kinks.
In principle, kink and anti-kink could annihilate with each other, 
but because of the integrability of the model, instead, they scatter or 
form bound states which are called breathers. Breathers have been extensively
studied and they are known, for example, to scatter with each other 
elastically, like kinks or anti-kinks.

In the inhomogeneous version of the sine-Gordon model, the coefficient in front
of the potential becomes a function of $x$. In this paper we consider a 
square well potential where the potential coefficient is one everywhere 
except in a 
small region of finite width where it takes a smaller value.
In physical applications the potential coefficient is usually determined by
one of the physical properties of the system. It can be the magnitude of a 
magnetic field or an other property of the considered material. 
The square well we have chosen
thus correspond to a system where those physical properties take two different 
values and where the transition between these values is very short compared
to the size of the inhomogeneity and the size of the sine-Gordon kink or 
breather.  

In a recent work \cite{PZsgw}, we have studied the scattering of a kink on 
the square well potential and now we investigate
how the breather scatters on the same well. The scattering of a breather
in inhomogeneous systems is not new. A few years ago F. Zhang\cite{Zhang} 
studied the scattering of the breather on a $1/\cosh^2$ potential. For very 
narrow well, the square well is very similar to the potential studied by 
him, while for a wide well, it is quite different, as in our case the 
non-integrability is generated by only two points, the edges of the well.
This means that breathers and kinks can freely propagate inside a well that 
is larger than their own size. This, as we will show, has a large influence on 
the scattering properties of the breather.

In what follows, we present a detailed study of the scattering of the 
breather on
a square well, analysing the dependence on all the parameters which influence
this scattering. As Zhang, we have observed 
several modes of scattering: transmission or reflection of the breather,
trapping of the breather and splitting of the breather into a kink and an 
anti-kink. We find a strong dependence of the scattering mode on the 
breather oscillation phase and then discuss the relative occurrence of these 
different modes when one varies the parameters of the model.

\section{Sine-Gordon Model with a Potential Well}
The sine-Gordon model with a square potential well is defined by the 
following Lagrangian
\begin{eqnarray}
{\cal L} = \int   {1\over 2} (f_t^2-f_x^2 - 2\,k(1-\alpha)\,(1-\cos(f))\, dx.
\label{SGlag}
\end{eqnarray}
where 
\begin{eqnarray}
\alpha &= a \qquad &-L/2 < x < L/2 \\
\alpha &= 0 \qquad  &\hbox{elsewhere}.
\end{eqnarray}
When $a$ is negative, the potential is thus a square well of width $L$ and 
depth $|a|$. The equation of motion is given by
\begin{equation}
f_{tt} -f_{xx} +k(1-\alpha)\,f = 0.
\label{SGsw}
\end{equation}

The scattering of a kink on a square well potential was studied in 
\cite{PZsgw} where it was shown that at small speeds, the kink becomes 
trapped by the well while at large speeds it goes through the well but looses
some energy through radiation and thus exits from the well with a speed 
smaller 
than the initial one. The speed above which the kink can escape from the well 
was called the critical velocity. 

For velocities in a few, very narrow, ranges of incoming 
speeds, just below the critical velocity, it was observed that the kink does 
neither go through the well nor get trapped in it, but instead bounces out of 
the well and returns to where it came from. Thus we have a reflection.

After studying the scattering of a kink on the square well, it is natural to 
ask what happens to a breather sent towards a similar well.

A breather which moves at speed $v$ is described by \cite{DJ}
\begin{equation}
f(x,t) = 4\, \hbox{atan}({\sin(\omega\,\sqrt{k}T) \sqrt{1-\omega^2}\over 
        \omega\cosh(\sqrt{1-\omega^2}\,\sqrt{k}X)})
\label{SGbreather}
\end{equation}
where $X = {x-vt\over \sqrt{1-v^2}}$ and $T = {t-vx \over \sqrt{1-v^2}}$.
This $f(x,t)$ is a solution of the pure sine-Gordon equation,  
{\it i.e.} (\ref{SGsw}) when $\alpha=0$.

The energy of the breather can be easily calculated and is given by 
$E = 16\, \sqrt{k} {\sqrt{(1-\omega^2)}\over \sqrt{(1-v^2)}}$. Notice that in 
our units, the energy of a kink is equal to $8$.
It is well known that the breather is a bound state of a kink and an anti-kink 
and thus the energy of the breather is less than the energy of a kink and an 
anti-kink infinitely separated ({\it i.e.} 16).

A stationary breather is a periodic function in time of period 
$T=2 \pi /(\omega k)$. For small values of $\omega$, the period of the 
breather is thus very large and, at the apex of the oscillations, the kink 
and the anti-kink are well separated. When $\omega$ is nearly 1, the period is
slightly larger than $2\pi/k$, the maximum amplitude for the breather is small
and the kink and anti-kink never really separate from each other.

The scattering of a breather on the well depends on several parameters:
the breather parameter $\omega$, the incoming speed $v$, the width $L$ and
the depth $a$ of the well as well as the phase of the breather when 
it hits the well. As the breather is an extended object, the scattering time 
cannot really be defined precisely and so it is difficult to accurately 
determine
the phase of the breather at the time of the scattering. Nevertheless it is 
straightforward to show that, in the well frame, the distance
travelled by the breather outside the well, {\it i.e.} when $k=1$, 
during one period of 
oscillation is equal to $d = 2\pi /(\omega \sqrt{1-v^2})$. 
To cover the full set 
of breather phases all we have to do is to put the breather initially at 
several positions within the range $[-x_1-d , -x_1]$ where $x_1$ must be 
sufficiently far away from the edge of the well so that, initially, 
the breather does not overlap with it.

To investigate the dependence of the scattering on the parameters $L$, $a$
$v$ and $\omega$ we have systematically scanned the full range of the breather
phase by varying the initial position of the breather in the range 
$[-x_1-d , -x_1]$ by step $dx=0.02$ (we have thus used over 300 values of the 
phase for each parameter set). We have then counted the number of times
each type of scattering has occurred and compared their relative occurrences.

Before we analyse the dependence of the scattering on these parameters
we first describe the different phenomena that we have observed, that is, 
transmission, trapping, backwards and forwards scattering as well as backwards
splitting.
Initially, we consider the case of $v=\omega$ so that the 
energy of the breather equals the energy of a kink and an anti-kink. This is 
the critical case where the kink and anti-kink cannot split outside 
the well. So, unless otherwise stated, in the discussions below it is 
assumed that $v=\omega$. Later we report on what happens when $v\ne\omega$.

\subsection{Breather Transmission}
When a breather is sent on a well, one would expect that if the incoming 
speed is large enough or if the well is small and shallow, the breather should
be able to go through it and emerge on the other side of the well. We have 
found that 
this is indeed the case, but the picture is more complicated than for the 
scattering of a kink on the well. We have found that the outcome of the 
scattering process is very sensitive to the phase of the breather. While at 
large speeds the breather  seems to always pass through the well, it is also
true that for most values of the well's width and depth, 
the soliton can also pass through the well for nearly any value of the speed if 
it has the right phase. 

\begin{figure}[htbp]
\unitlength1cm \hfil
\begin{picture}(16,16)
 \epsfxsize=8cm \put(0,8){\epsffile{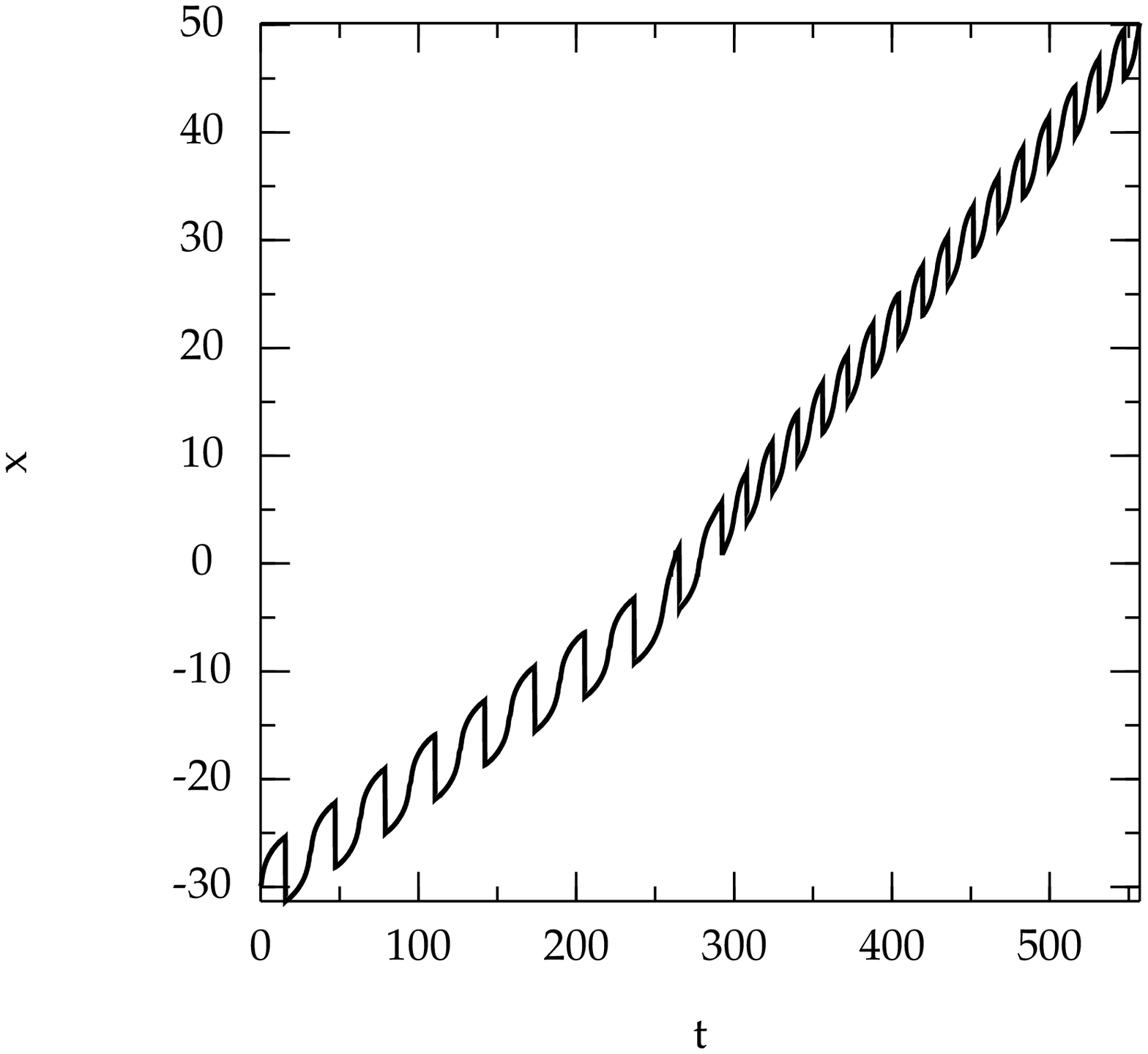}}
 \epsfxsize=8cm \put(8,8){\epsffile{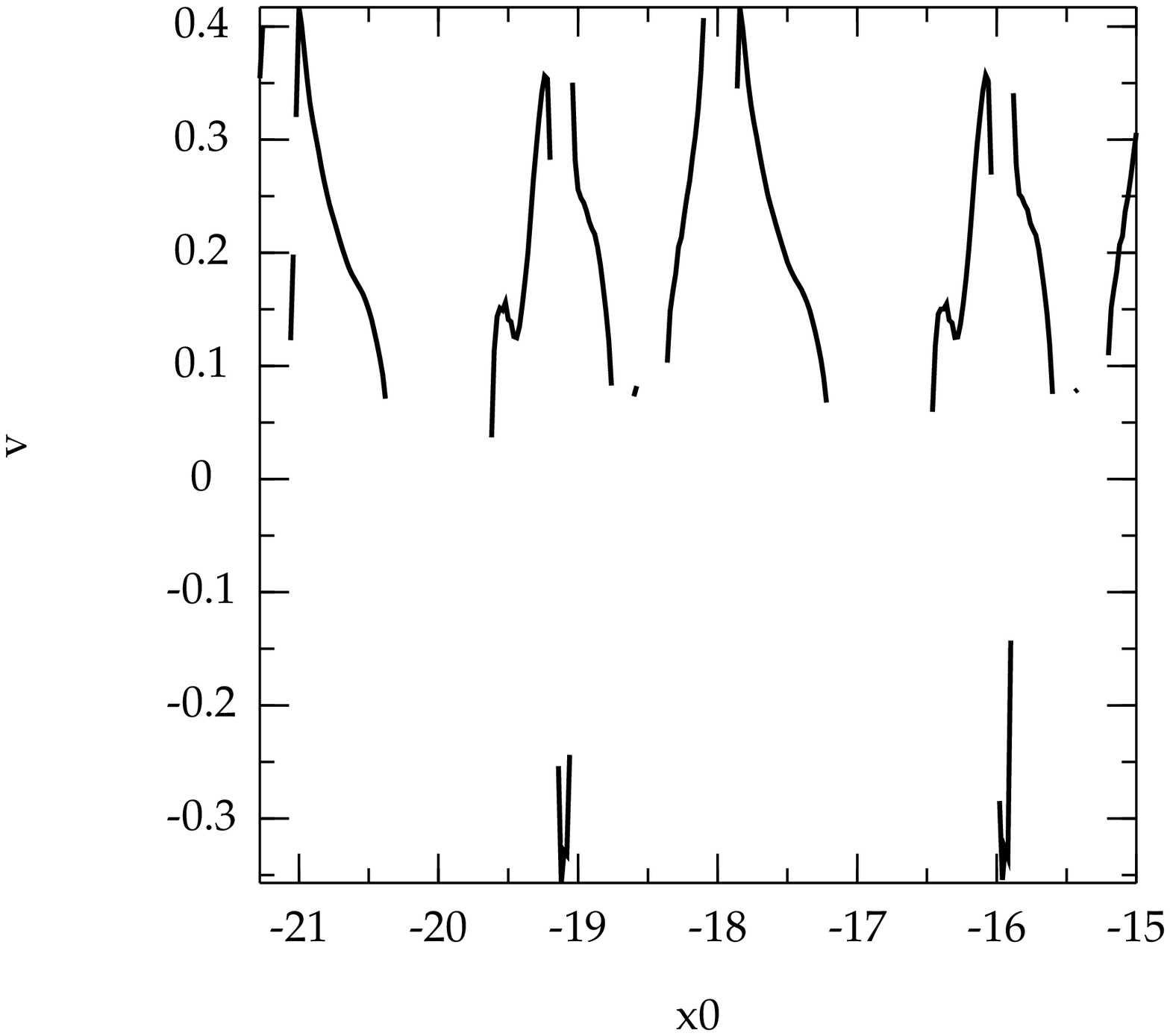}}
 \epsfxsize=8cm \put(0,0){\epsffile{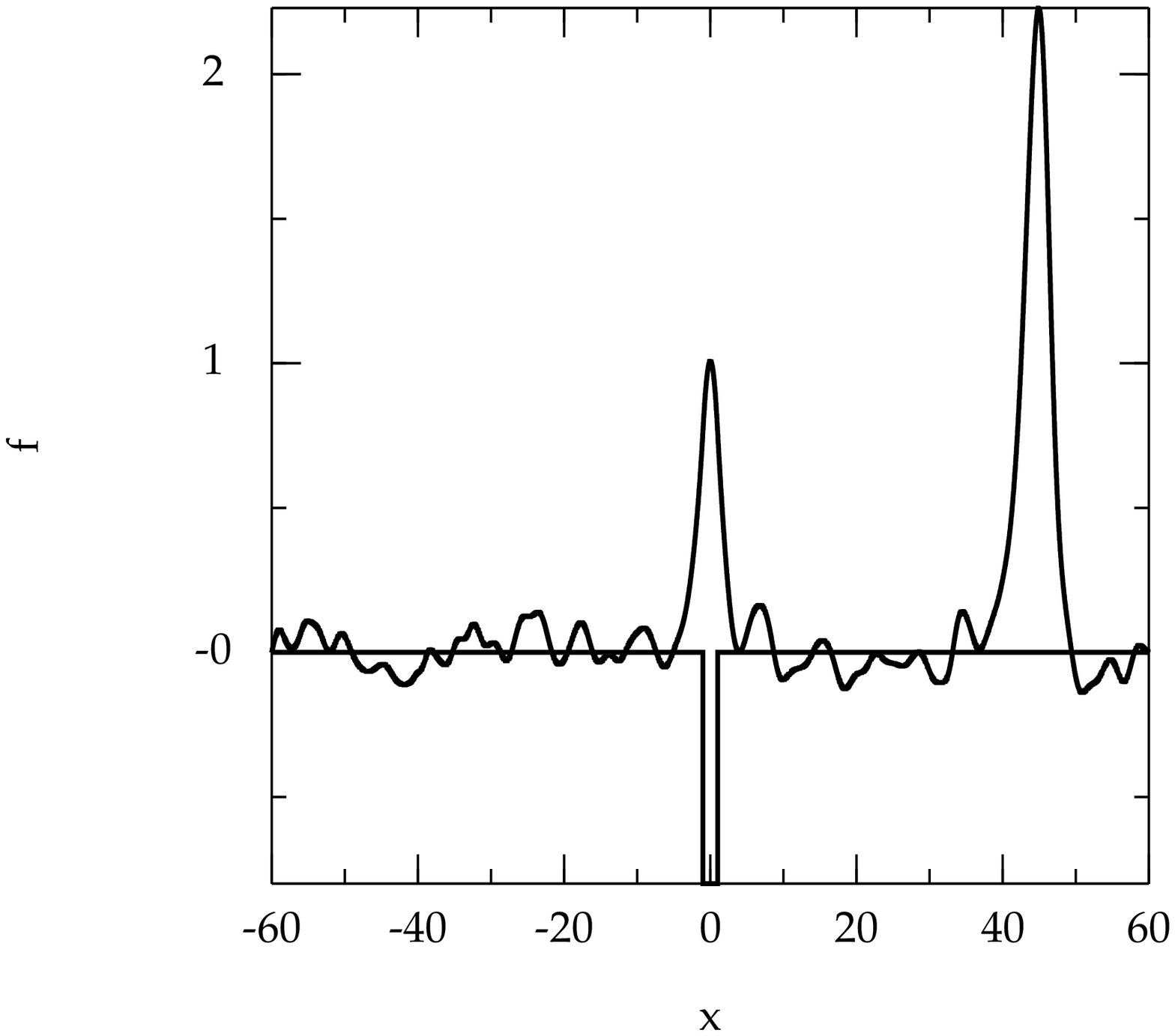}}
\put(4,8){a}
\put(12,8){b}
\put(4,0){c}
\end{picture}
\caption{ a) Breather position for a  
well with $L=2$, $a=-0.2$, $v=\omega=0.1$ and $x_0=-29.92$.
 b) Breather outgoing speed for a well with $L=10$, $a=-0.2$, $v=\omega=0.1$.
The gaps correspond to other types of scattering.
c) A solution profile for $L=2$, $a=-0.8$, 
$v=\omega=0.3$ and $x_0=-15.5$. The breather is split into two: one is ejected 
from the well while the other one is trapped inside it.
}
\label{fig1}
\end{figure}

It is important to stress that the scattering is always inelastic. When the 
breather crosses the well it always radiates some energy away. The energy
of the outgoing breather is thus always smaller than the initial energy.
The radiated energy can come from two different sources:  
the kinetic energy of the breather or its internal energy.
We have observed that indeed both energies can decrease but more surprisingly
we have also seen that one can increase while the other one decreases,
with the total, of course, decreasing.
In particular, we have observed that sometimes the kinetic energy of the 
breather increase during the scattering. In these cases, the well thus acts 
like a slingshot. This is illustrated on Figure \ref{fig1}a where we present 
the 
position of the maximum of the energy density as a function of time. The 
oscillations are caused by the vibration of the breather and we clearly see
that the speed of the breather jumps from $v=0.1$, before the scattering, to 
$v=0.17$ afterwards. The period of oscillation also changes from $T=62.517$
to $T=31.87$, corresponding to $\omega=0.1$ and $\omega=0.194$, respectively.

In Figure \ref{fig1}b we present the outgoing speed of the breather as a 
function the breather's initial position ({\it i.e.} its phase)  for the 
case $L=10$, $a=-0.2$ and $v=\omega=0.1$. Notice that in this case the 
outgoing speed of the breather is nearly always larger than it initial speed.
This is a generic feature we have observed for $v=\omega=0.1$. When we took
$v=\omega=0.3$ we notioced that the outgoing speed tends to 
oscillate typically in the range $[0.2, 0.4]$.
Looking at figure \ref{fig1}b, notice also the small amount of backwards 
scattering described below.

More surprisingly we have seen a few cases of a breather being split in two 
breathers (Fig \ref{fig1}c). 
One is ejected from the well while another one remains trapped 
inside the well. This leads us to another observed scattering outcome: the
trapping of the breather.

\subsection{Breather Trapping}
When a breather scatters on a well, it can become trapped in it. As we 
will see later, this occurs more often when the well is deep. Once trapped 
the breather is actually unstable and it slowly radiates away its energy.
This happens because of the perturbation introduced by the well. 

As is well 
known \cite{LLMQ}, in the linear limit, a square well always has at least one 
vibration mode:
\begin{equation}
f(x)\,=\,\left\{ \begin{array}{ll} 
         A \sin(\delta) \exp(-k\,\hbox{ctan}(\delta) (x+L/2))  & x < -L/2\\
         A \sin(K (x+L/2) + \delta))   & -L/2 \le x \le L/2 \\ 
         A \sin(\delta) \exp(-k\,\hbox{ctan}(\delta)(L/2-x)) & L/2 < x\\ 
      \end{array},\right.
\end{equation}
where $\delta = \arcsin(\sqrt{1+a k L^2/4})$ and
$K = \sqrt{-a k (1 + a k L^2/4)}$. The period of oscillation of this vibration 
is given by
\begin{equation}
T = \frac{2 \pi}{ \sqrt{k(1 - a^2 k \frac{L^2}{4})}}
\end{equation}
and we see that, like the breather, the period is always larger than $2\pi$.
Note that the ground state obtained in the linear limit and the breather 
are very similar and can be thought of as sine-Gordon vibrations in the well 
derived in 2 different limits. The breathers are vacuum excitations of the 
sine-Gordon model. When confined in a large well they can have any width 
for as long as they fit inside the well. The amplitude of the breather is 
linked to its size and frequency. The linear vibrations in the well, on the 
other hand, have a fixed size and period, but their amplitude is arbitrary for 
as 
long as they remain small. In a relatively narrow well, the breather does not 
really fit inside the well and so becomes a large amplitude linear 
vibration. 

The sine-Gordon vibrations in the well always decay by radiating away some 
energy: the {\it linear 
vibrations} radiate because of the nonlinerarity of the sine-Gordon 
equation while the
breather radiates because of the perturbation introduced by the well.

As stated in the previous section we have sometimes observed a breather going 
through the well but leaving a relatively large oscillation in it. In 
figure \ref{fig1}c the amplitude of oscillation for the excitation inside 
the well 
and the outgoing breather are $1.7$ and $2.7$ respectively
(this is not visible in Figure \ref{fig1}c simply because the two 
oscillations are out 
of phase and we have chosen to plot the figure at the time when 
both excitations are relatively 
large). The formation of a double breather is very rare though; out of 
100000 or so simulations that we have performed, we have only seen the creation 
of a double breather a few times in the regions of parameters corresponding 
to the transition between a breather trapping 
and a breather transmission.

\subsection{Breather Backwards Scattering}
The breather can sometimes bounce out of the well. As we will show 
later, this tends to occur mostly at small speeds and in a narrow 
and shallow well, but unlike what happens
for the scattering for the kink, this scattering mode is observed for large
ranges of the parameters values. As for the forwards scattering, the breather
can be ejected backwards from the well with a speed larger than it had 
initially. This is seen in Figure \ref{fig1}b in two narrow regions of $x_0$
where the outgoing speed of the breather is negative.

\begin{figure}[htbp]
\unitlength1cm \hfil
\begin{picture}(8,8)
 \epsfxsize=8cm \put(0,0){\epsffile{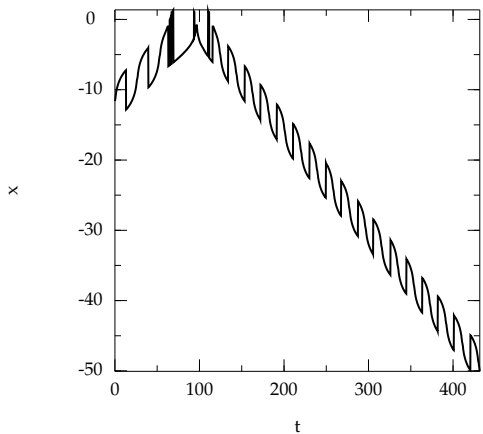}}
\end{picture}
\caption{Position of the breather in the case of a 
well centred  on $x=0$ for $L=2$, $a=-0.2$, $v=\omega=0.12$ and $x_0=-11.6$. 
}
\label{fig2}
\end{figure}

In Figure \ref{fig2} we present the time evolution of the position of the 
breather during a backward scattering for the case $L=2$, $a=-0.2$, 
$v=\omega=0.12$ and $x_0=-11.6$. In this case the outgoing velocity is 
$v_{out} = -0.144$.

\subsection{Breather Splitting}
The most interesting phenomenon we have observed when scattering a breather 
on the well is the spitting of the breather into a kink and an anti-kink pair.
As the energy of a breather or kink decreases when they fall into the well,
the excess or energy can be used to split the breather into a kink anti-kink 
pair. One of 
them remains inside the well while the other escapes from it and moves 
backwards 
or forwards. Initially, we expected this phenomenon to occur only
in a very narrow region of the parameter space. It turns out, however, that, 
together 
with  the transmission of a breather through the well, this is the most 
frequent outcome of the scattering process. The kink or anti-kink can be 
ejected from the well
in either direction but, as will be shown later, the forward scattering is 
more frequent, especially for shallow and wide wells. 

As the energy of a breather of frequency $\omega$ and speed $v$, 
outside the well, is $E_{br} = 16 \sqrt{(1-\omega^2)}/\sqrt{1-v^2}$
and the energy of a kink trapped inside the well, assumed to be large 
enough to contain the kink, together with an anti-kink outside it
is $E_{kak} =8(1+\sqrt{1-a})$ we can easily evaluate the critical 
speed below which the splitting is impossible:
\begin{equation}
v_c= \bigl(1-\frac{4(1-\omega^2)}{(1+\sqrt{1-a})^2} \bigr)^{\frac{1}{2}}.
\end{equation}
When $v\le\omega$, we find that $v > v_c$ and so we see that the splitting 
is always possible from an energetic point of view. 

Notice that the trapping of a kink and the ejection of an anti-kink is 
equivalent to the trapping of an anti-kink and the ejection of a kink. 
To see this, we observe that if we multiply $f(x,t)$ by $-1$, a kink becomes 
an anti-kink and vice-versa, while the breather becomes a breather with the 
opposite phase. So any solution with a trapped kink and an ejected anti-kink
can be transformed into a solution with a trapped anti-kink and an ejected kink
by changing the phase of the breather by 180 degrees.

The scattering of the breather on the well is inelastic and generates some 
radiation waves, so not all the trapping 
energy is transferred to the ejected kink (anti-kink). We have also 
also observed that 
after the scattering, the kink and anti-kink wobble a little, radiating 
some energy away (the sine-Gordon kink does not have genuine vibration modes
\cite{Kevrekidis}).
Moreover, the trapped kink (anti-kink) moves back and forth inside the well 
even when the well is quite narrow.

\begin{figure}[htbp]
\unitlength1cm \hfil
\begin{picture}(16,8)
 \epsfxsize=8cm \put(0,0){\epsffile{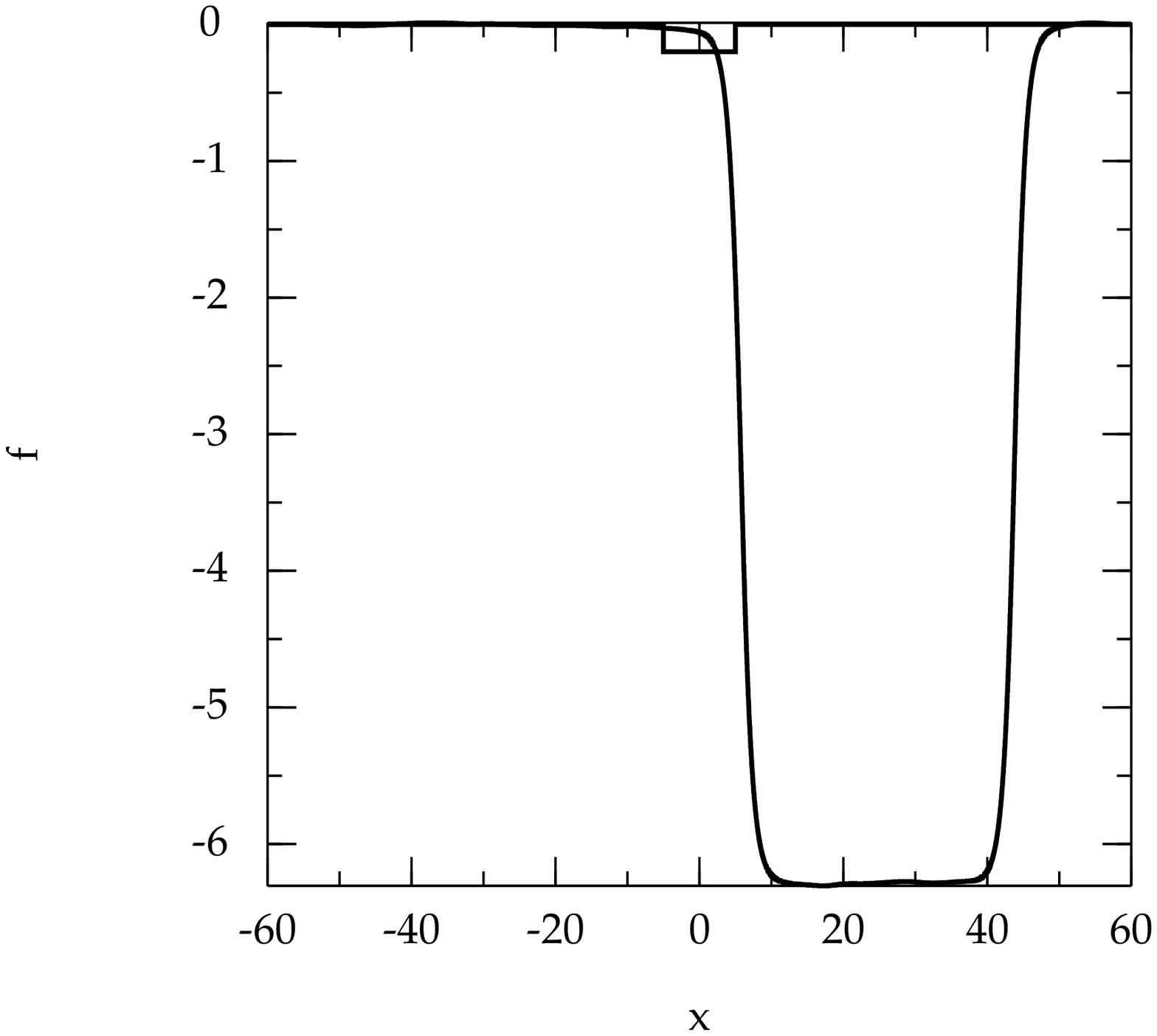}}
 \epsfxsize=8cm \put(8,0){\epsffile{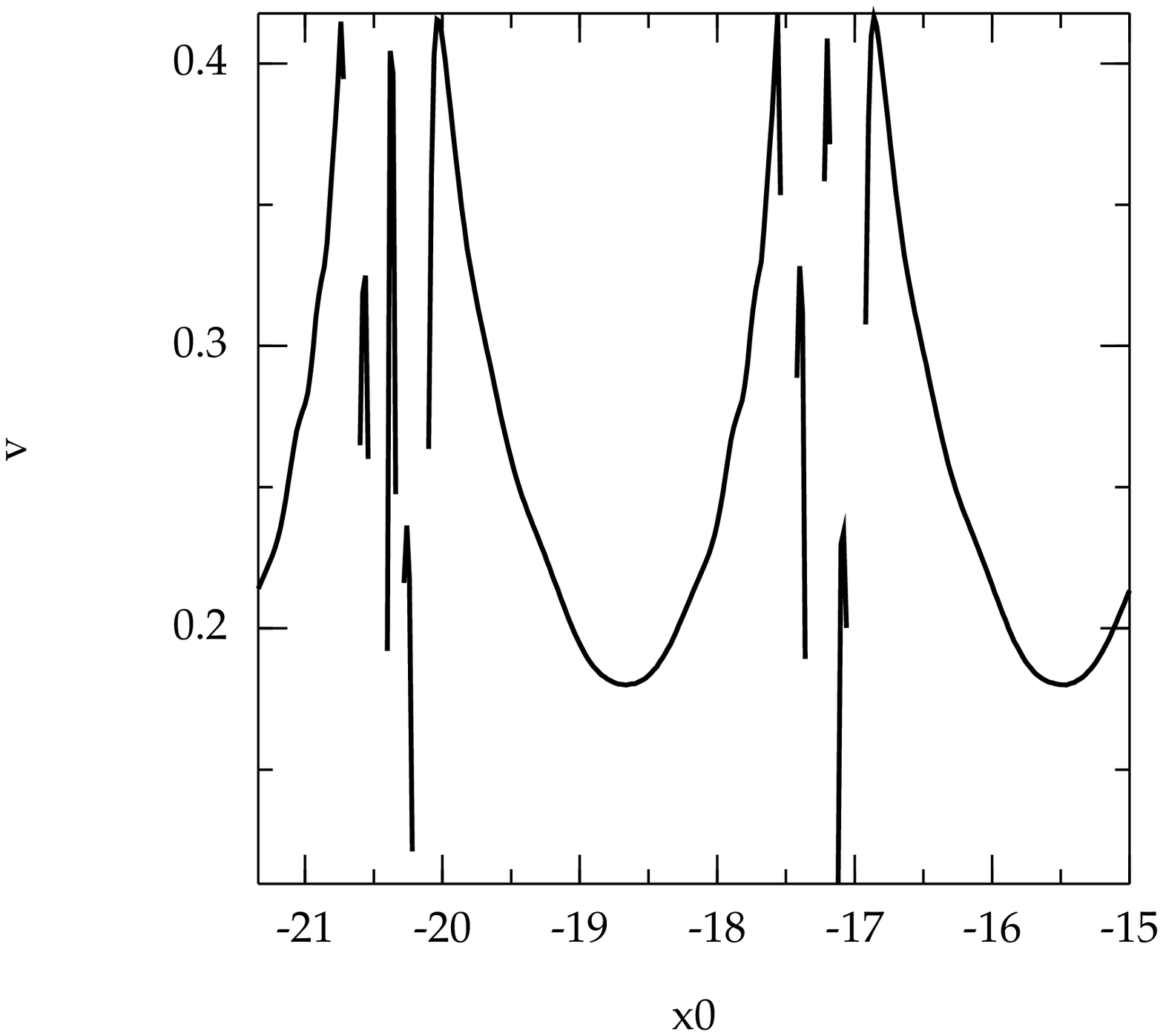}}
\put(4,0){a}
\put(12,0){b}
\end{picture}
\caption{Split breather scattering with $L=10$, $a=-0.2$, $v=\omega=0.14$. 
a) Profile after the scattering for $x_0=-17.12$. 
b) Outgoing speed of the kink or anti-kink (the gaps correspond to the 
regions of forwards scattering). 
}
\label{fig3}
\end{figure}

In Figure \ref{fig3} we show the profile of a split breather after its 
scattering on a well with $L=10$, $a=-0.2$, $v=\omega=0.14$ and $x_0=-17.12$.
The outgoing speed of the ejected kink in this case is $v_{kink}=0.109537$.
The speed of the ejected kink or anti-kink varies with the phase of the 
breather. This is shown on Figure \ref{fig3}b where we present the outgoing 
speed  of the kink after the scattering. Note that the speed of the outgoing 
kink (anti-kink) is nearly always larger than the incoming speed of the 
breather. This is true in most cases but it varies a little with the depth and 
the width of the well.

\section{Parameter Dependence}
\begin{figure}[htbp]
\unitlength1cm \hfil
\begin{picture}(16,8)
 \epsfxsize=8cm \put(0,0){\epsffile{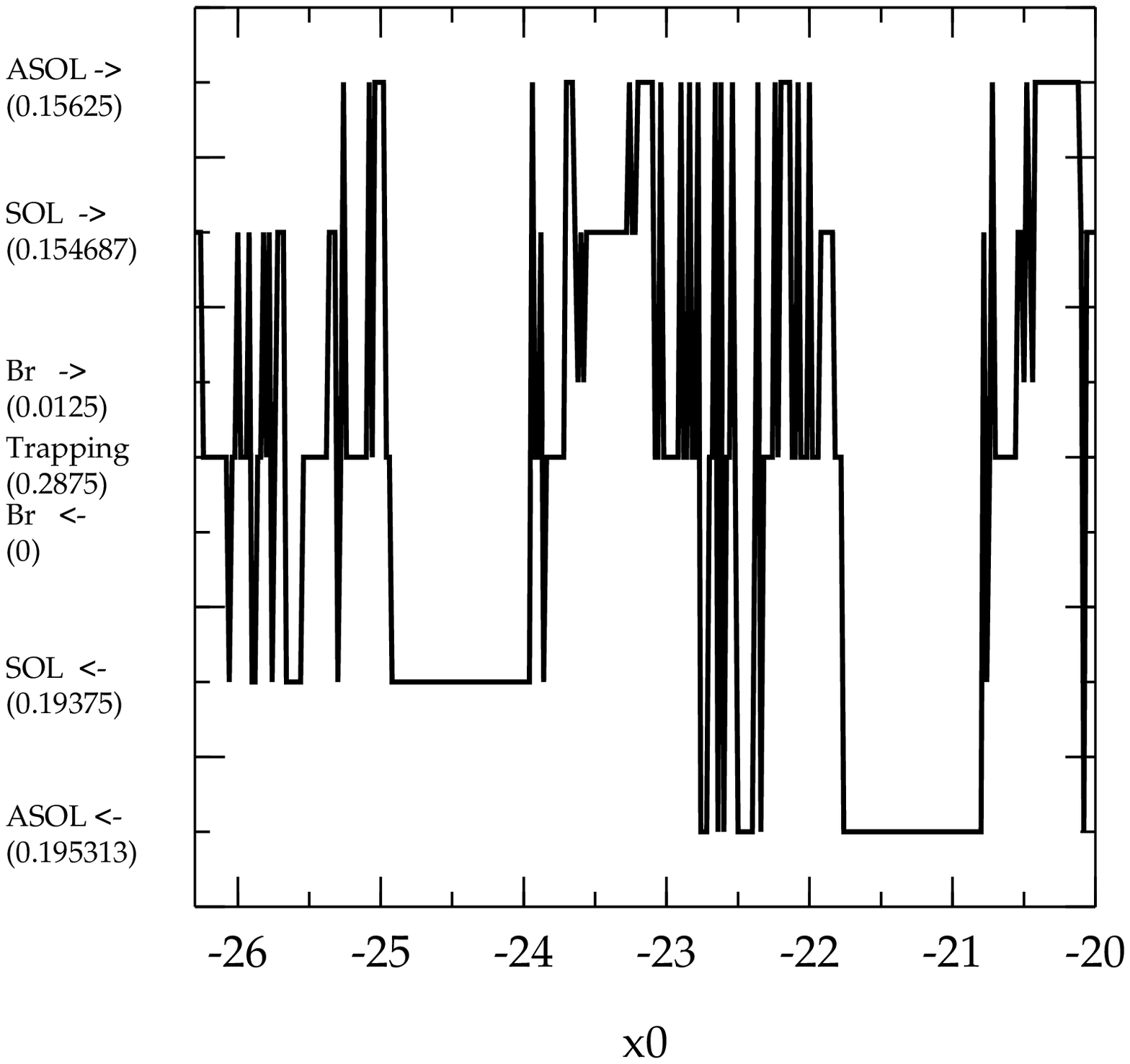}}
 \epsfxsize=8cm \put(8,0){\epsffile{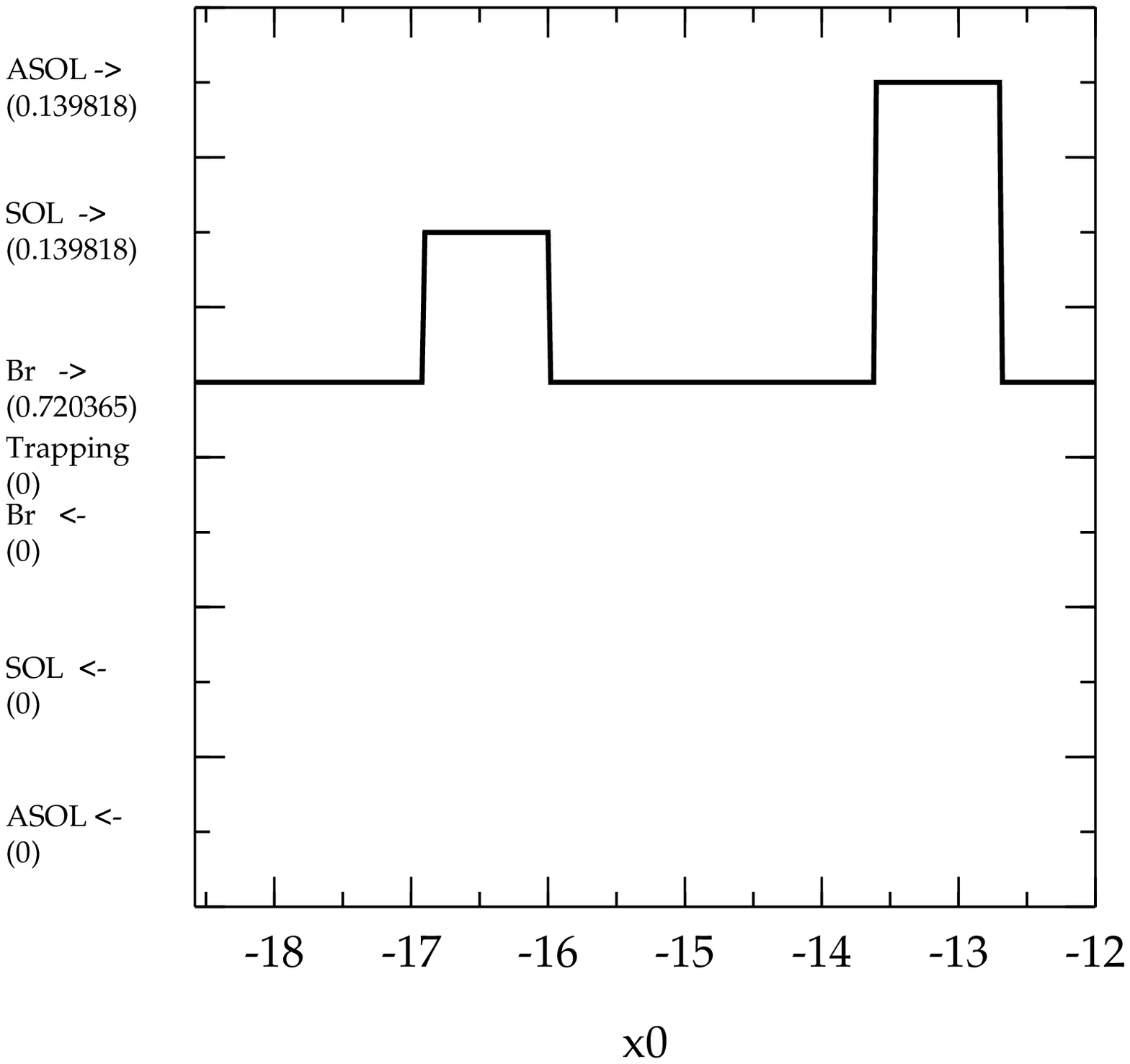}}
\put(4,0){a}
\put(12,0){b}
\end{picture}
\caption{ Phase dependence of the scattering modes for a breather 
a) $a=0.5$, $L=20$, $v=\omega=0.1$;
b) $a=0.2$, $L=2.4$, $v=\omega=0.3$
}
\label{fig4}
\end{figure}

The most important parameter in determinning the properties of the 
scattering of the breather on the 
square well is the phase of the breather. In many instances, the scattering 
outcome is very sensitive to its value. This is shown in figure \ref{fig4} 
where we present the scattering mode as a function of the breather phase 
({\it i.e.} the breather initial position ) in two extreme cases. 

In Figure \ref{fig4}a, we have $a=0.5$, $L=20$ and $v = \omega=0.1$, and 
we see that the 
scattering mode is very sensitive to the breather phase. This is very common 
for small values of $v$ and $\omega$, especially when most scattering 
modes can occur. The numbers in parenthesis below the mode names correspond to
the total fraction of these modes for the full range of the phase.

In Figure \ref{fig4}b, we have  $a=0.2$, $L=2.4$ and $v=\omega=0.3$, and now 
the scattering 
is very regular: the breather splits into a kink or an anti-kink in two well
defined phase regions and it scatters forwards in other cases. We have 
observed this type of pattern especially when the speed is large. Our 
explanation for this is that when the breather moves slowly, it has 
time to oscillate several times inside the well. Its vibration phase 
thus changes relatively rapidly during the scattering process, 
leading to different scattering modes. 
At large speeds, on the other hands, the breather mostly goes through the well
and scatters with the phase it has at that time. This fits well with the 
observations we have made when studying the scattering of a baby-Skyrme 
soliton on a square 
well\cite{PZ_BSW} where at low speeds, the soliton oscillates several times 
inside the well before  emerging from it on one side of the well or the other.

\subsection{Dependence on $v$ and $\omega$}
By letting $v=\omega$, for the breather, its energy is exactly equal to the
energy of a kink and an anti-kink infinitely separated. Thus this is the 
critical value for being able to generate a pair of kink and anti-kink 
outside the well and as such it is a natural choice of parameters to study the
scattering modes of the breather. We will consider what happens when  
$v\ne\omega$ in a later section.

\begin{figure}[htbp]
\unitlength1cm \hfil
\begin{picture}(16,8)
 \epsfxsize=9cm \put(0,0){\epsffile{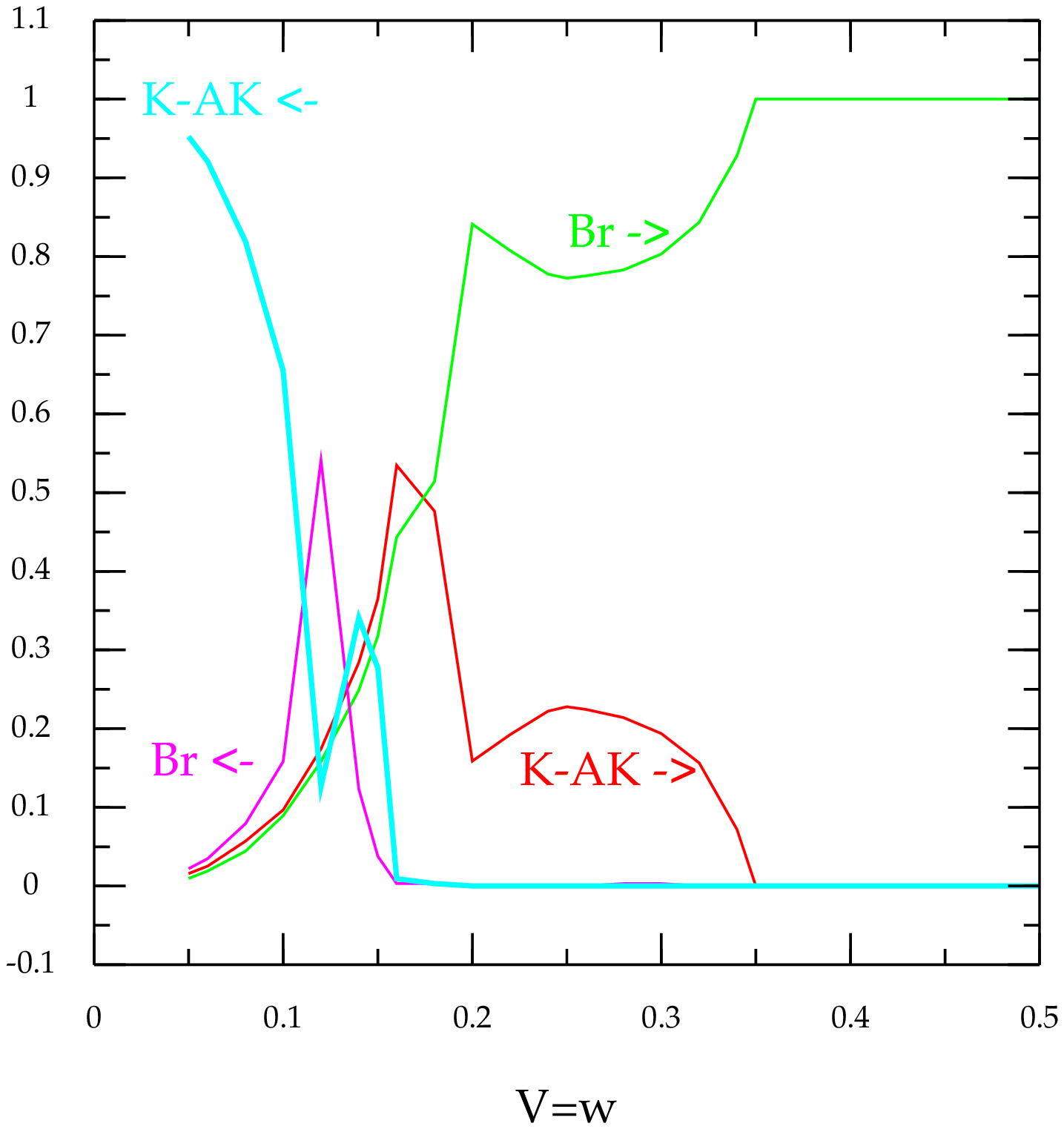}}
 \epsfxsize=9cm \put(8,0){\epsffile{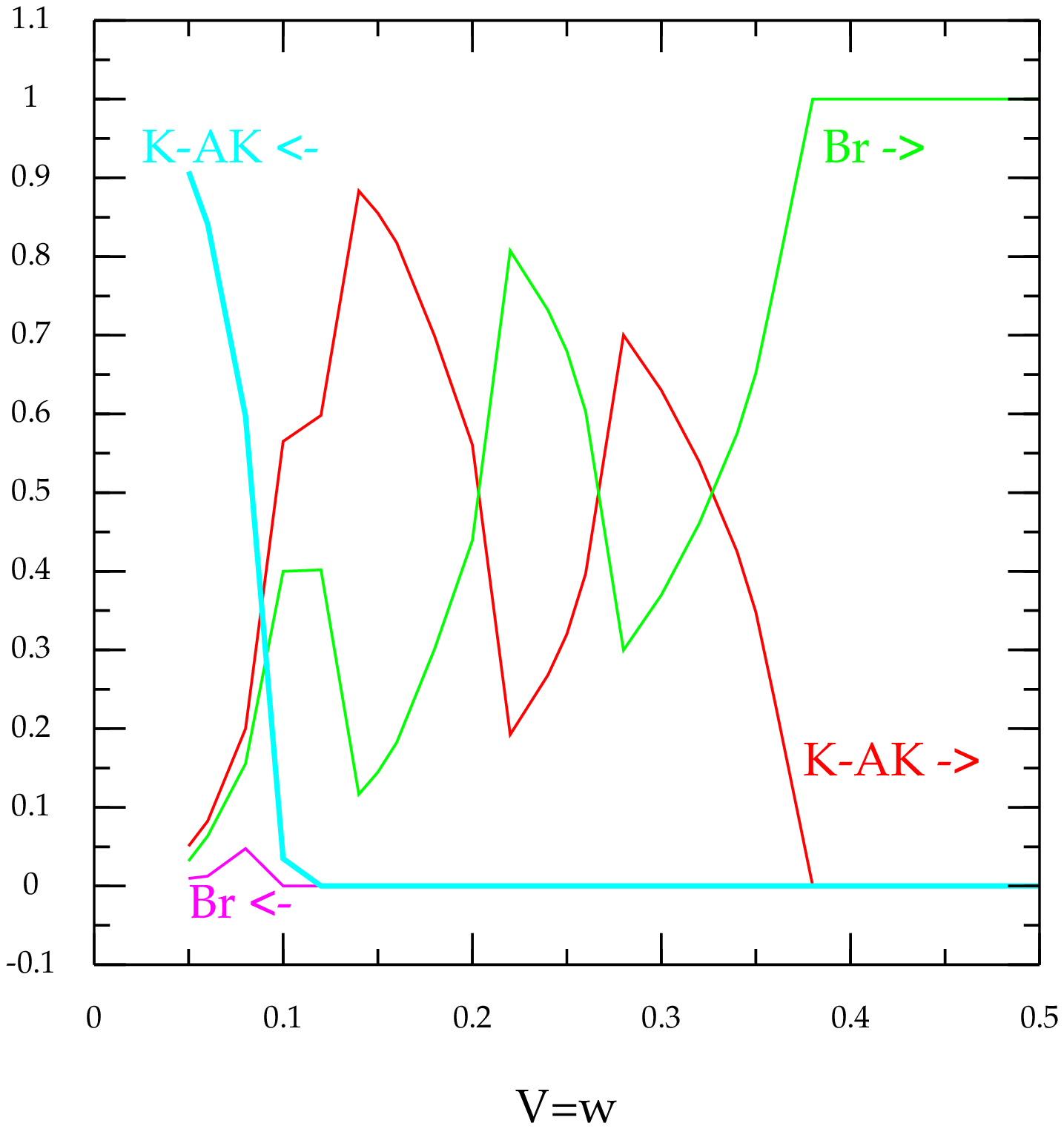}}
\put(4,0){a}
\put(12,0){b}
\end{picture}
\caption{Scattering mode frequencies as a function of $v=\omega$.
a) $a=0.2$, $L=2$ ;
b) $a=0.2$, $L=10$. 
"$Br-<$" : forwards breather scattering.
"$Br<-$" : backwards breather scattering.
"$K-AK->$" : trapped kink (anti-kink)  and forwards anti-kink (kink).
"$K-AK->$" : trapped kink (anti-kink)  and backward anti-kink (kink).
}
\label{fig5}
\end{figure}

In figure \ref{fig5}, we present the relative occurrence of the different
scattering mode as a function of the speed  $v=\omega$ for two different wells:
$a=0.2$, $L=2$ (fig \ref{fig5}a) and $a=0.2$, $L=10$ (fig \ref{fig5}b).
For small values of $v$ and $\omega$, the breather are dominantly splits into a 
kink and a anti-kink with one of them trapped inside the well while the other
escapes backwards. This is a general feature at low speeds when the well is 
narrow and reasonably deep.
Looking at movies that we have made of several scattering of this type, 
we have always observed the following: when the breather hits the well, the 
kink 
(anti-kink) falls into the well and get trapped. Because of the narrowness 
of the well, the anti-kink (kink) is neither able to fall inside it nor 
to push
the kink (anti-kink) outside the well. It has thus no other choice but to
bounce on the trapped kink (anti-kink). 
If the initial speed is increased sufficiently, the second anti-kink (kink)
has enough energy to push the trapped kink (anti-kink) at least partially 
out of the well. This can then result in a forwards or backwards 
scattering as well as in the splitting of the kink where a kink or an anti-kink
is ejected forwards.

At large speeds, there are only two scattering modes: forwards transmission 
or forwards splitting. Note the oscillations 
between these two modes in figure \ref{fig5}b. At large $v$, there is always
only one scattering mode, the transmission of the breather. This can be easily 
explained by the fact that the breather has enough energy to cross the well
quickly without being affected much by it.

\subsection{Dependence on the well width $L$}
\begin{figure}[htbp]
\unitlength1cm \hfil
\begin{picture}(16,8)
 \epsfxsize=9cm \put(0,0){\epsffile{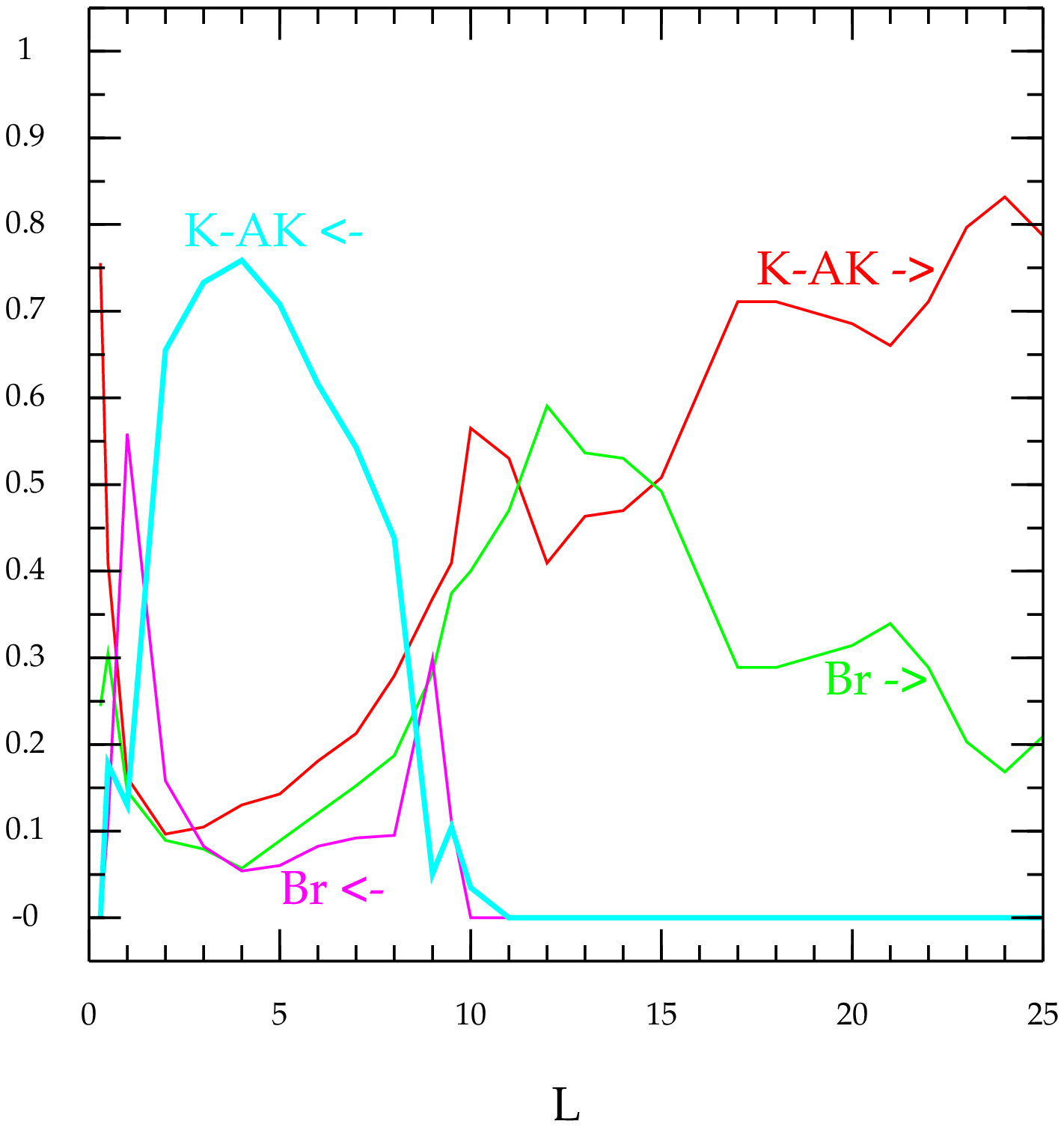}}
 \epsfxsize=9cm \put(8,0){\epsffile{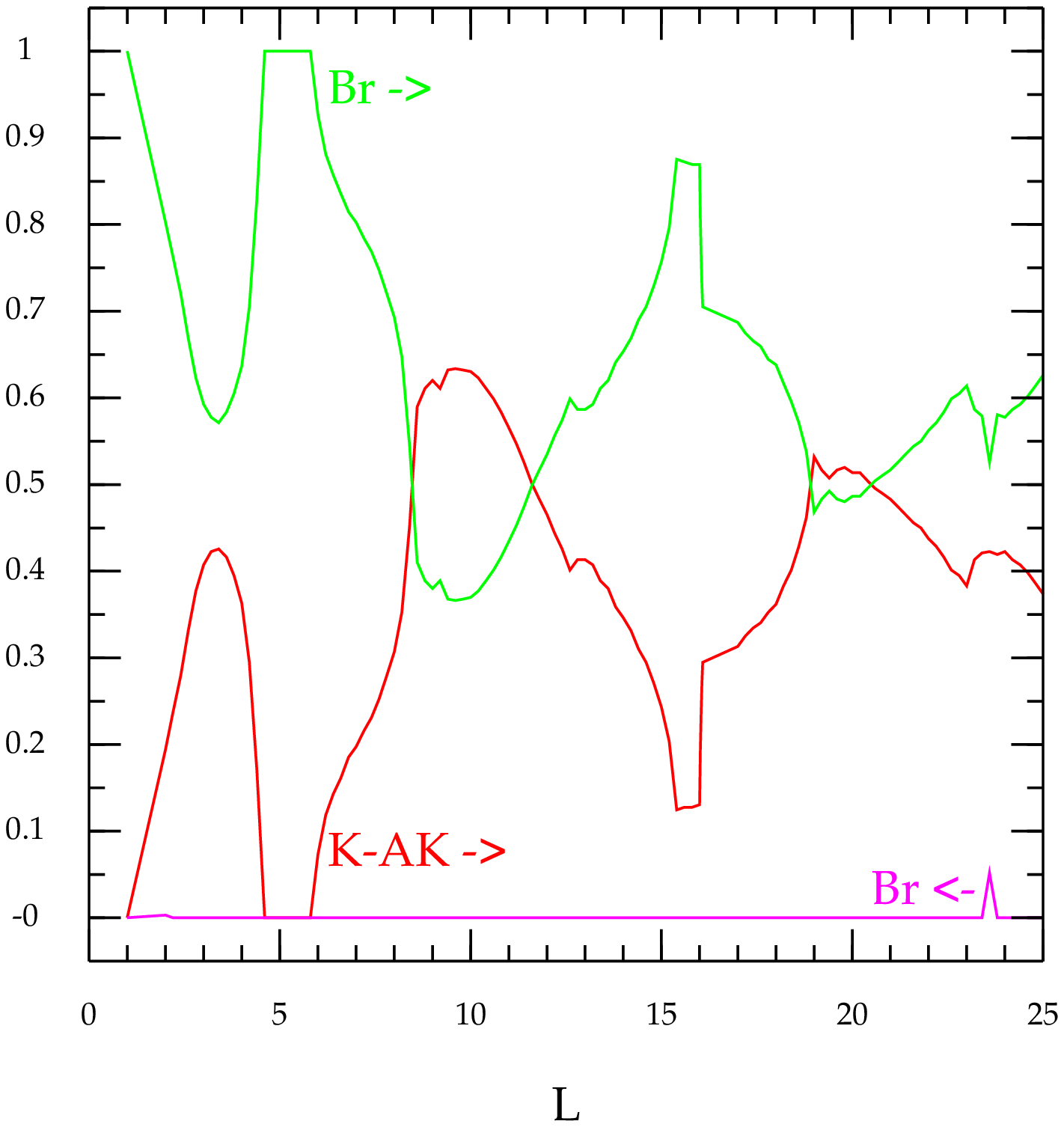}}
\put(4,0){a}
\put(12,0){b}
\end{picture}
\caption{Scattering mode frequencies as a function of $L$.
a) $a=0.2$, $v=\omega=0.1$ ;
b) $a=0.2$, $v=\omega=0.3$
}
\label{fig6}
\end{figure}

The width of the well also affects the scattering of the breather, as
is shown in figure \ref{fig6}a for $a=0.2$, $v=\omega=0.1$ 
and in figure \ref{fig6}b for the cases $a=0.2$, $v=\omega=0.3$.

When $v=\omega=0.1$ and for very narrow wells, the dominant scattering mode 
is the forwards splitting of the breather. The well is so narrow that the 
second kink has enough energy to push the first one out of the well.
When the well is wider than 2 the dominant mode is the backwards
splitting of the breather as explained in the previous section. 

Once the well is larger than 10, only 2 scattering modes are rellevant 
in the two 
cases presented: the forwards transmission and the forwards splitting. 
What we have observed when the well is very large is that when the breather 
enters the well its splits into a kink and an anti-kink pair. 
For a reason we have not understood yet, whichever of the kink or anti-kink
is in front happens to have more energy than its partner. 
The kink and the anti-kink then move 
forwards, slowly increasing the distance separating them until  one of them 
hits the other side of the well. As it has more energy, it manages to 
climb out of the well and to escape forwards while the second soliton, 
with less energy, remains trapped inside the well.

The phase of the breather when it falls into the well determines which of the 
kink or the anti-kink, into which the breather has divided, is at the front 
and eventually escapes from the well.
In the intermediate region between the two modes, there is a succession of 
narrow windows where the breather goes through the well, separated by 
windows where the kink or the anti-kink, into which the breather split, 
escapes from the well. 

\subsection{Dependence on the well depth $a$}

The depth of the well plays a major role in the breather scattering as
this is the parameter that determines the binding energy and the size of a 
breather or a kink inside the well.

\begin{figure}[htbp]
\unitlength1cm \hfil
\begin{picture}(16,16)
 \epsfxsize=9cm \put(0,8){\epsffile{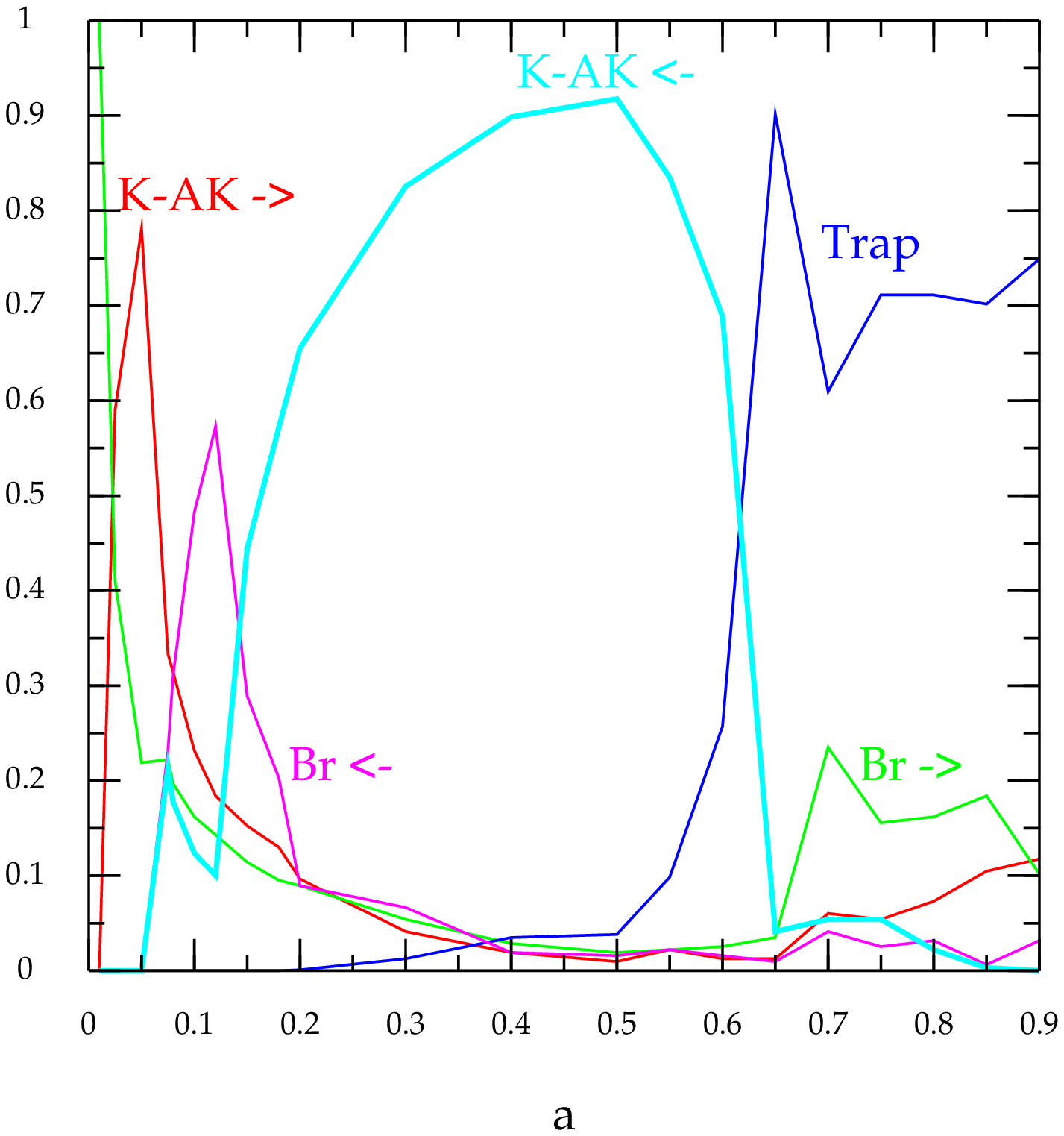}}
 \epsfxsize=9cm \put(8,8){\epsffile{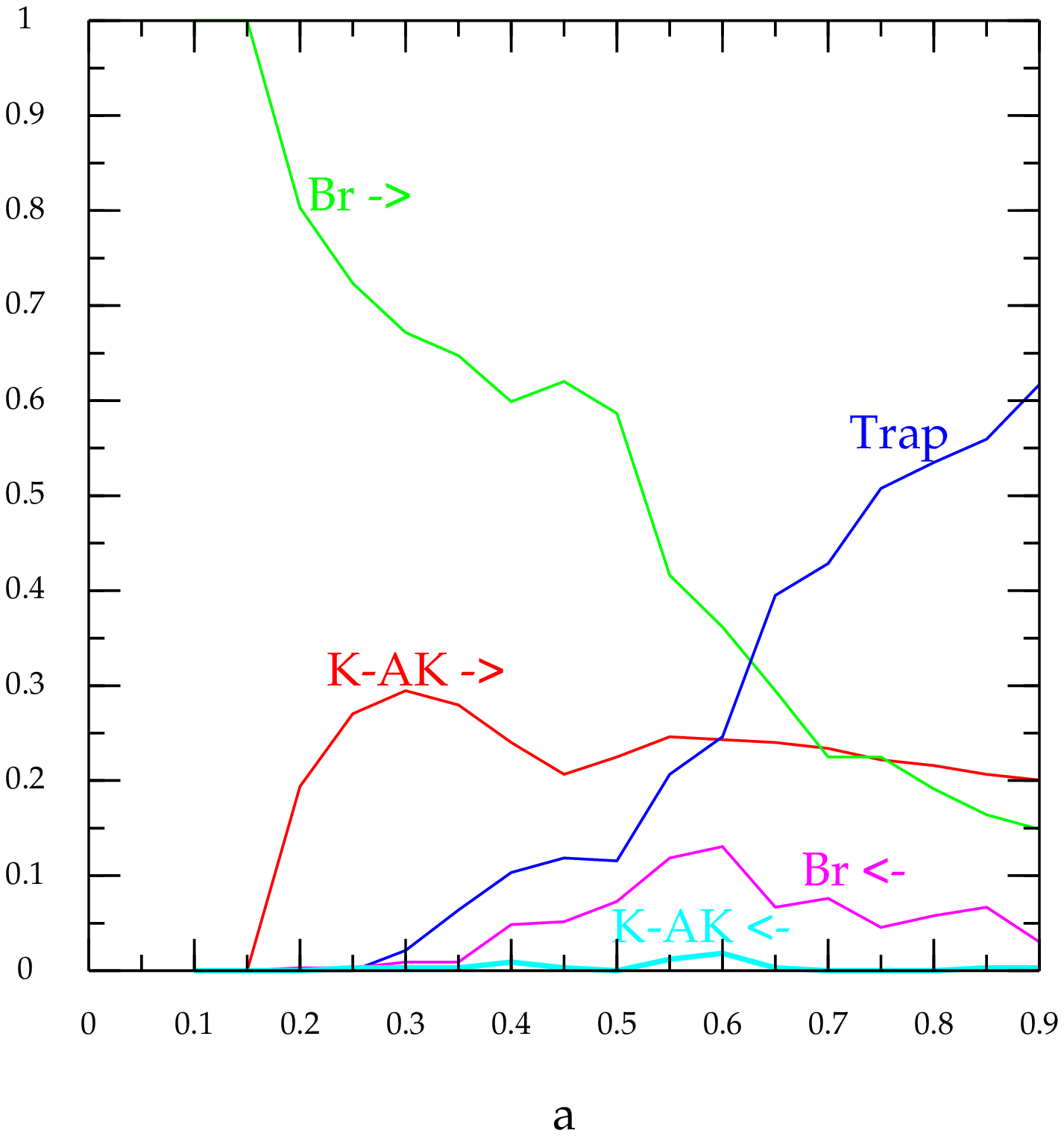}}
 \epsfxsize=9cm \put(0,0){\epsffile{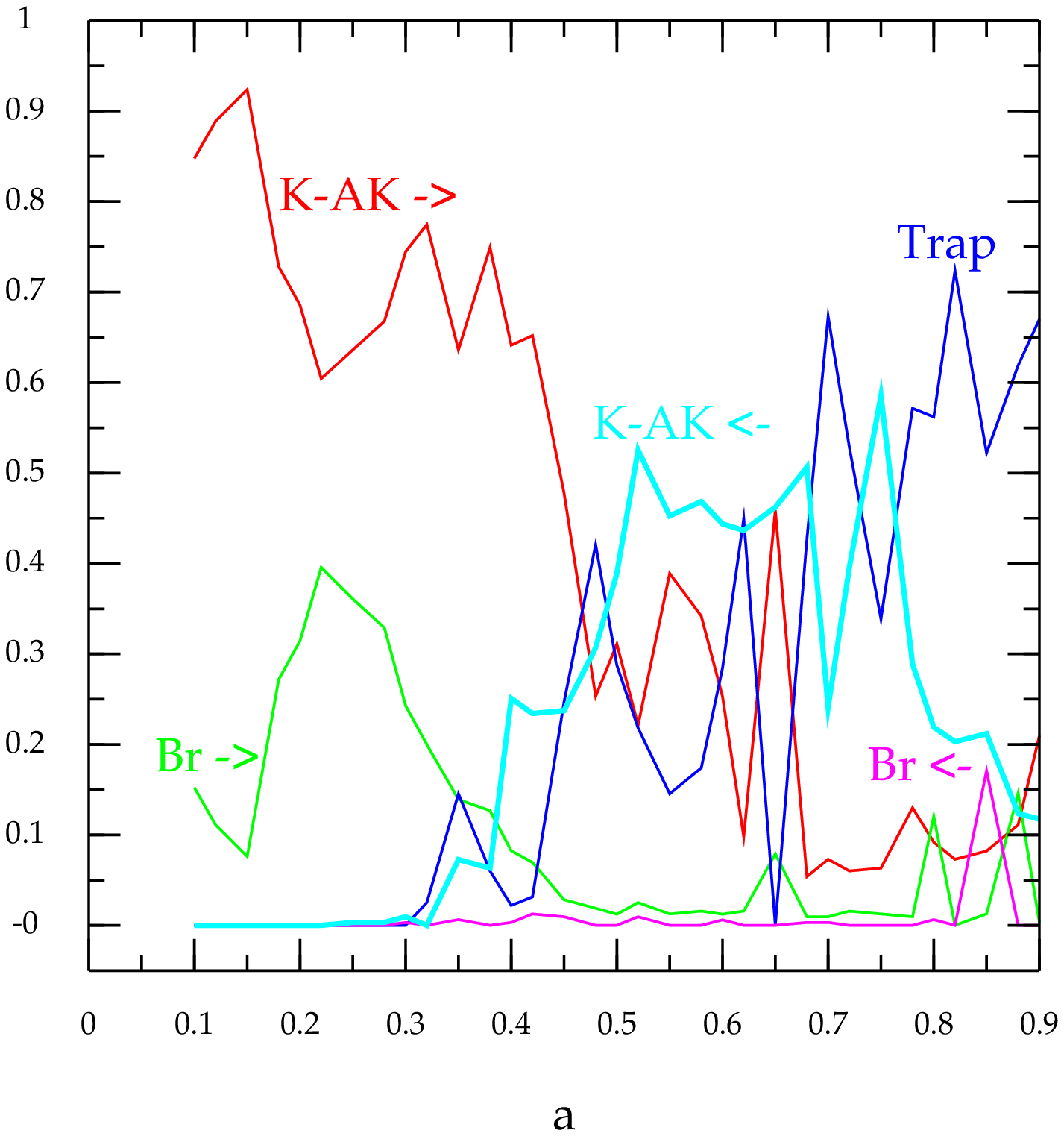}}
 \epsfxsize=9cm \put(8,0){\epsffile{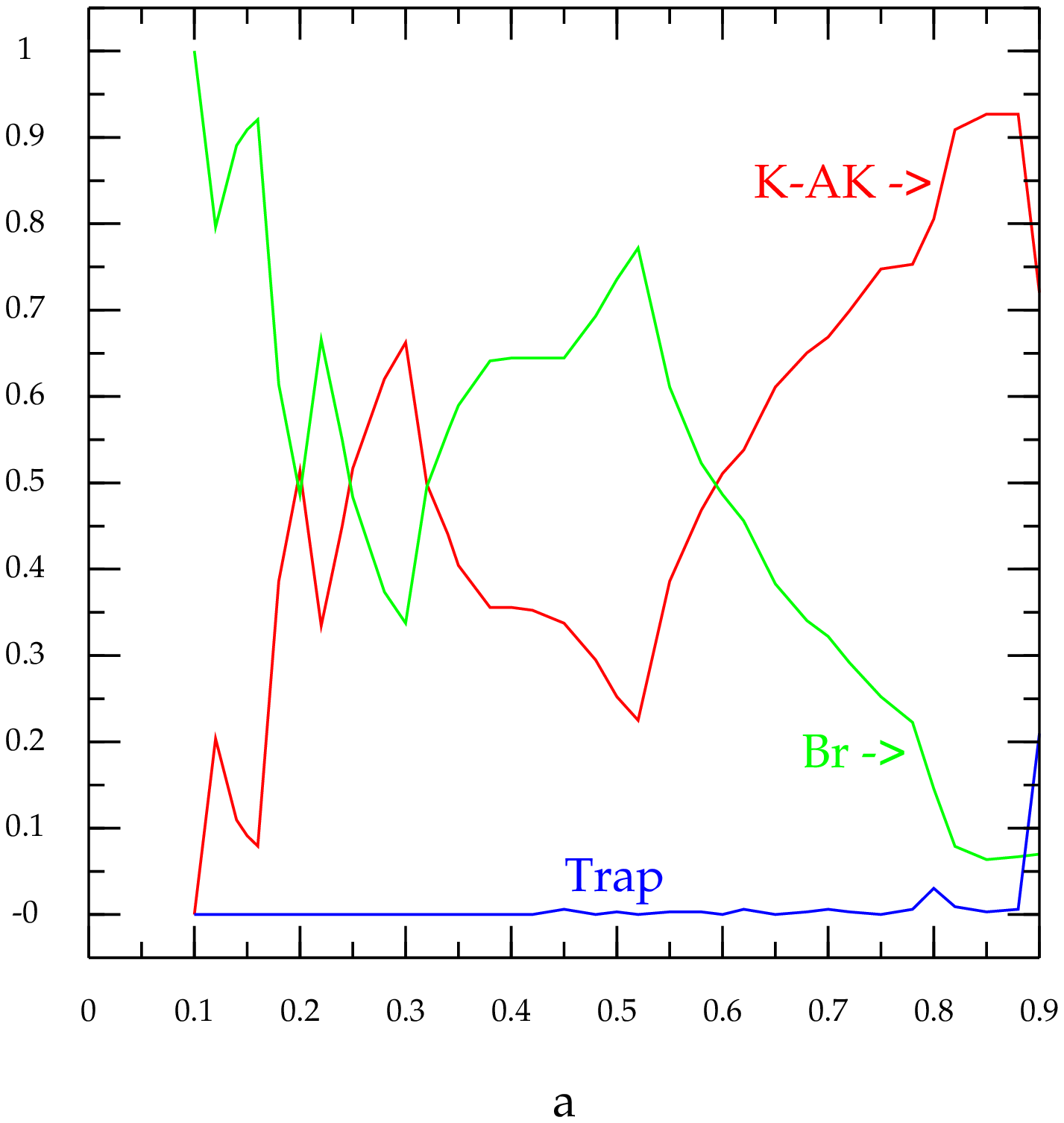}}
\put(4,8){a}
\put(12,8){b}
\put(4,0){c}
\put(12,0){d}
\end{picture}
\caption{Scattering mode frequencies as a function of $a$.
a) $L=2$, $v=\omega=0.1$;
b) $L=2$, $v=\omega=0.3$;
c) $L=20$, $v=\omega=0.1$;
d) $L=20$, $v=\omega=0.3$;
}
\label{fig7}
\end{figure}

As shown in figure \ref{fig7} a, 
for a narrow well ($L=2$)  and at a small speed ($v=\omega=0.1$),
the scattering modes vary greatly with the depth of the well.
For very shallow wells ($a<0.02$), the dominant mode is the forwards
transmission: the breather hardly sees the well at all. 
For marginally deeper wells ($0.02<a<0.15$), the dominant modes are the
forwards splitting and then the backwards scattering. For deeper
wells ($0.15<a<0.6$), as explained in a previous section, the backwards
splitting dominates. Then when ($a > 0.6$) the well is so deep that the 
breather is usually trapped by the well.

When the speed is increased to $v=\omega=0.3$, see figure \ref{fig7}b which 
looks very much like a stretched out version of figure \ref{fig7}a, 
except for the trapping curve. In this
case, the breather has enough energy to go through the well quickly and 
the dominant mode is the forwards transmission until about $a > 0.6$ where
the trapping dominates.

For wide wells, the picture changes significantly. The as explained in the 
previous section, the breather nearly always splits into a kink anti-kink
pair inside the well. For the case presented in figure \ref{fig7}c,
$L=20$ and $v=\omega=0.1$, the parameters are such that the
forwards splitting is dominant for ($a < 0.45$). For deeper wells,
the scattering mode changes very rapidly as $a$ increases, but overall, 
the breather trapping dominates.

When the speed is increased to $v=\omega=0.3$, see figure \ref{fig7}d, the two
dominant modes are the forwards splitting and the forwards transmission. 
Trapping occurs rarely, only for very deep wells.
In this case, the breather has enough energy to interact with the well quickly
and, at least in part, escape from it.

\section{Varying $v$ and $\omega$ separately.}
So far we have looked only at the scattering of a breather on a well 
for the special case $v=\omega$, that is when the breather has exactly the 
same energy as an infinitely separated pair of a kink and an anti-kink. 
This critical 
case is particularly interesting, but it is also very interesting to investigate
what happens when $v$ and $\omega$ differ.

\begin{figure}[htbp]
\unitlength1cm \hfil
\begin{picture}(16,16)
 \epsfxsize=9cm \put(0,8){\epsffile{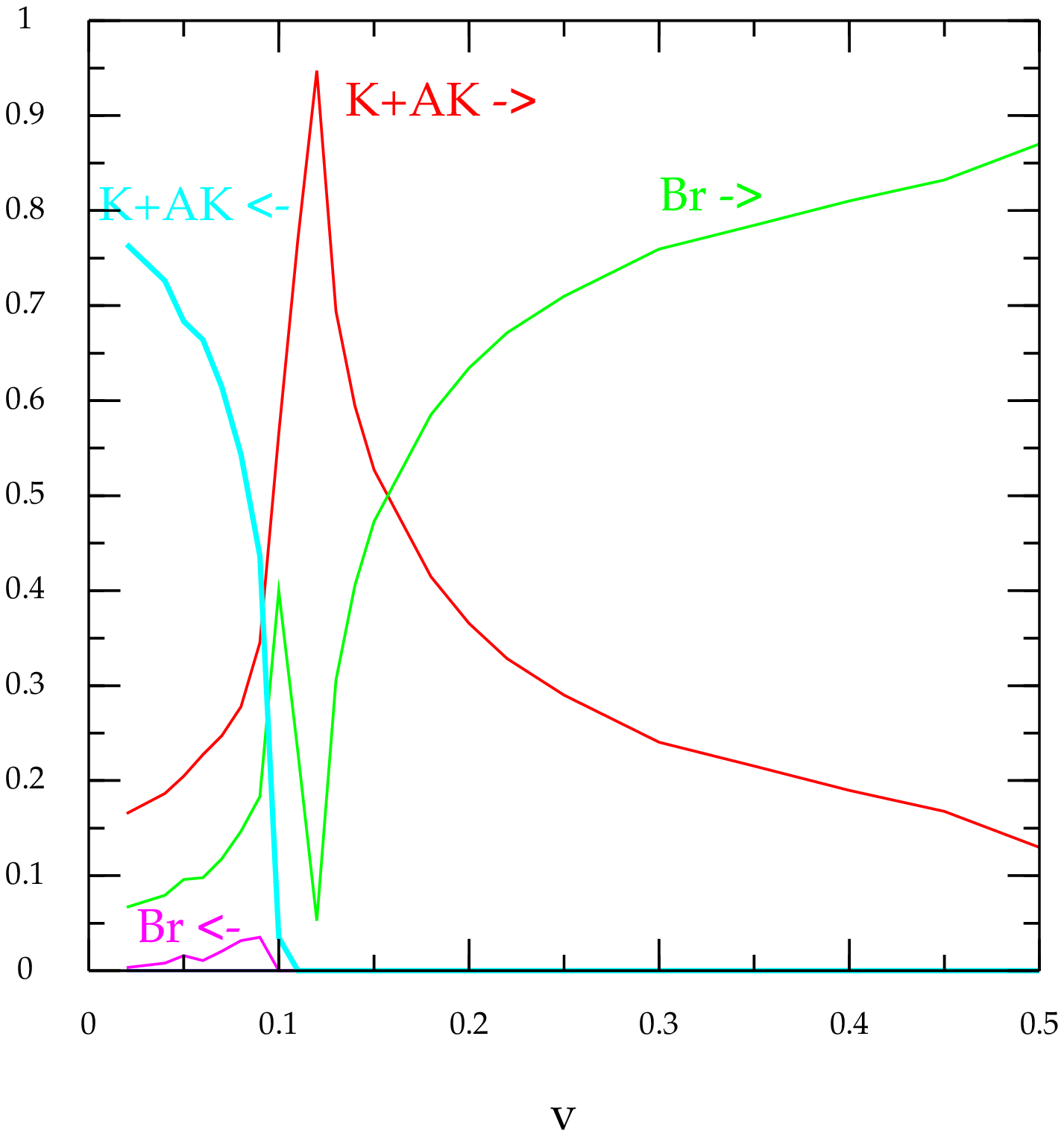}}
 \epsfxsize=9cm \put(8,8){\epsffile{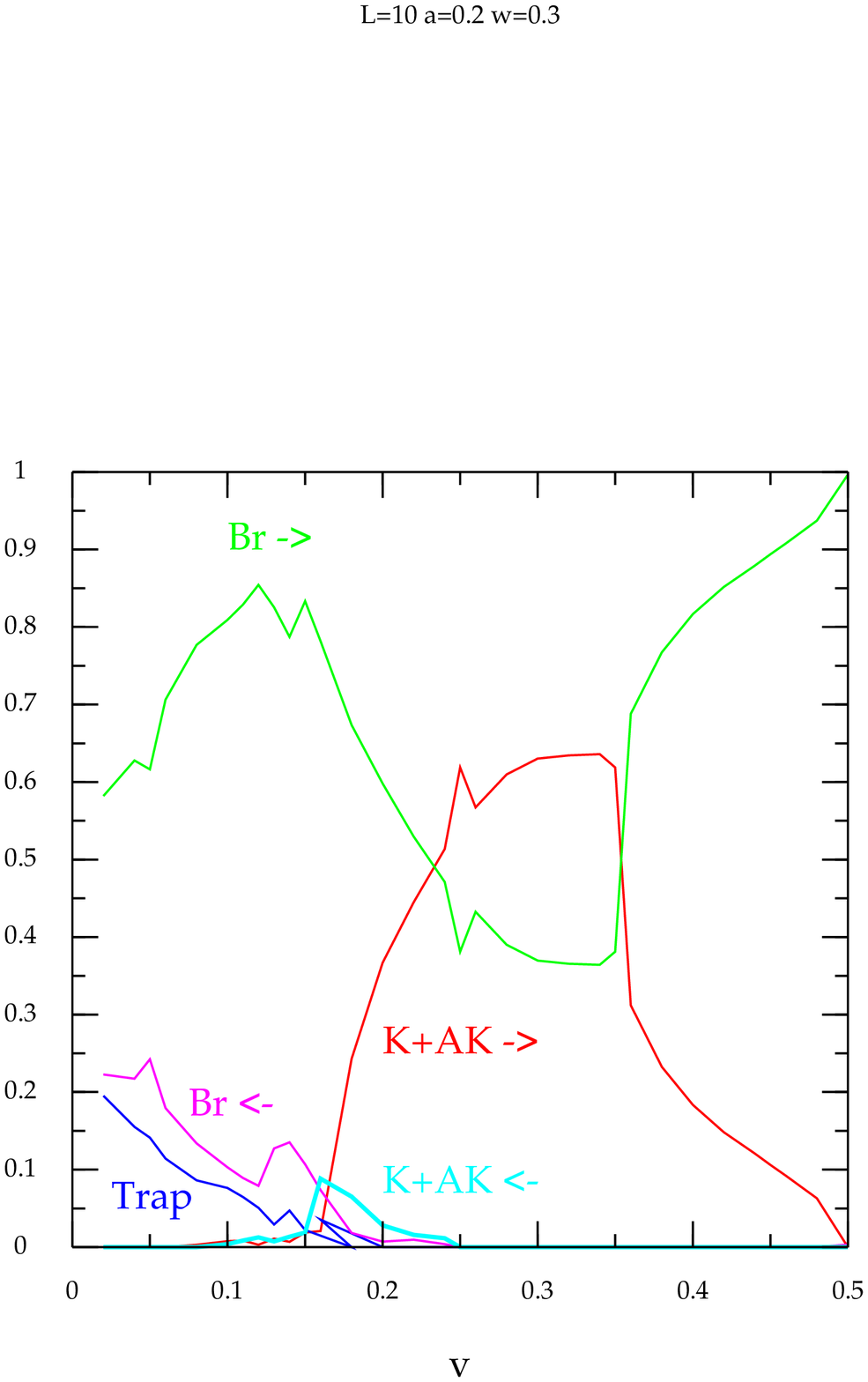}}
 \epsfxsize=9cm \put(0,0){\epsffile{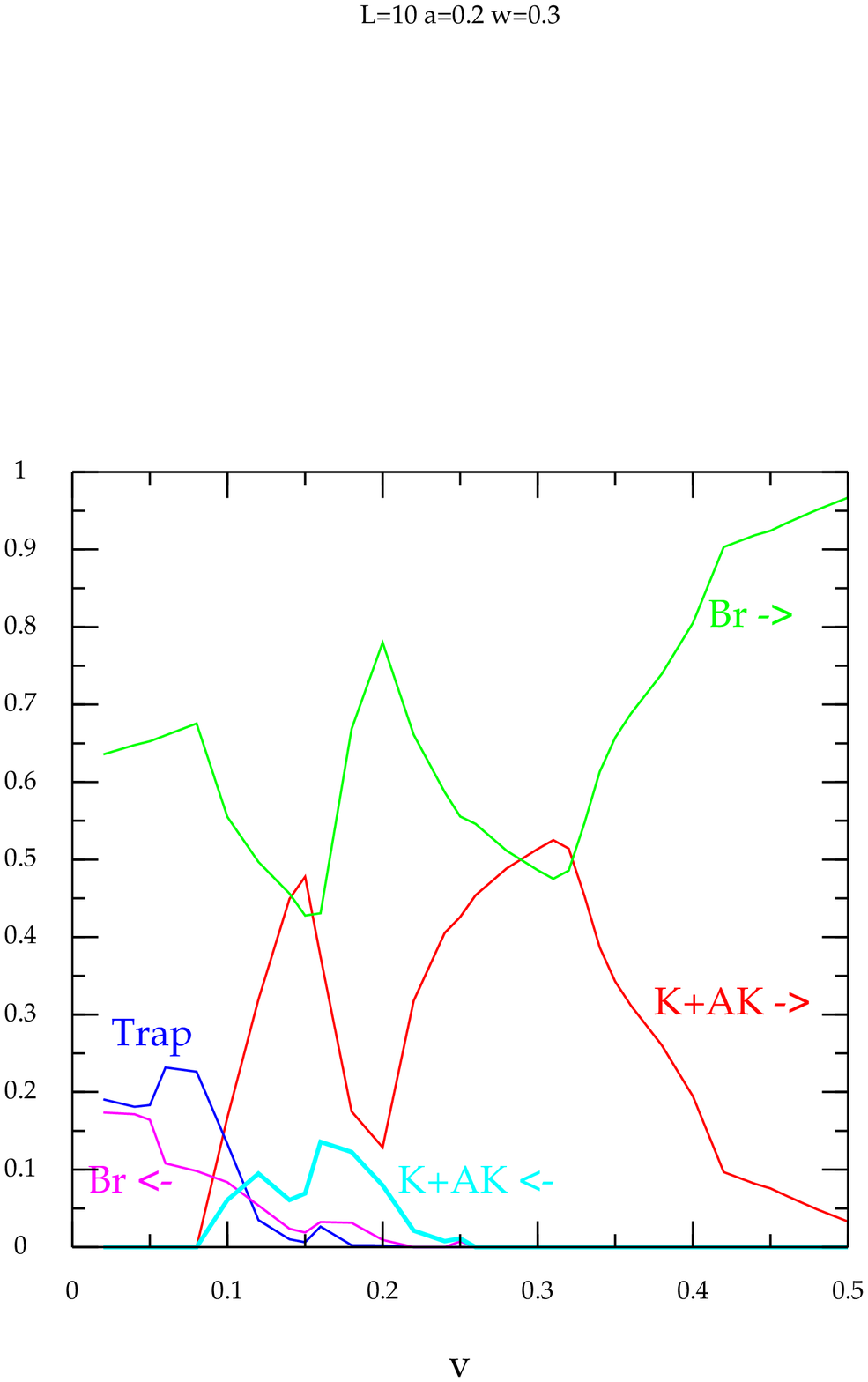}}
\put(4,8){a}
\put(12,8){b}
\put(4,0){c}
\end{picture}
\caption{Scattering mode frequencies as a function of $v$ for $a=0.2$ and
a) $L=10$, $\omega=0.1$;
b) $L=10$, $\omega=0.3$;
c) $L=20$, $\omega=0.3$;
}
\label{fig8}
\end{figure}

The results are shown of Fig \ref{fig8} where we have taken $a=0.2$ for
the depth of the well and we have considered a well of widths $L=10$ and $L=20$
for $\omega=0.1$ and $\omega=0.3$.

Looking at the figures we note 
that when $v >> \omega$, the breather is dominantly 
going through the well. In the three cases we have considered, 
the largest rates 
of kink anti-kink splitting occur in the region where $v \approx \omega$, the 
actual maximum being reached when $v$ is slightly larger than $\omega$.
 
When $v$ is smaller than $\omega$, all the familiar modes, 
{\it i.e.} trapping, backward scattering of the 
breather and backward kink anti-kink splitting, can all occur but at a 
relatively small rate. This can be explained by the fact that, having less 
kinetic energy, the breather is more likely to bounce on the far side of the 
well and, if the phase is right, come out of the well.

\section{Conclusions}
In this paper, we have shown that the scattering of a breather on a square well
exhibits very interesting phenomena. The breaking of the sine-Gordon 
integrability due to this inhomogeneity leads to scattering modes that are 
forbidden in an integrable model. In particular the sine-Gordon breather can 
be split into a kink and an anti-kink which move both forwards and backwards. 
Somewhat 
surprisingly, this scattering mode is genuine and sometimes, depending on the 
parameters of the model, the dominant one. 

Another surprising phenomenon seen in the scattering is that the well can 
accelerate the breather. This is possible because the internal 
energy of the breather can be partly 
converted into kinetic energy. This acceleration can occur for the 
forwards as well as the backwards motion. 

The parameter dependence of the scattering data is quite complicated.
It is also very sensitive to the phase of the breather when it 
collides with the well. 
Overall, we have observed that at high energies, the breather tends to 
scatter forwards or to split forwards into a kink/anti-kink pair while at
low energies on the other hand, all the scattering modes can take place.   

Given our observations, it would be interesting to find out if one can 
reverse the scattering process and create a breather by the scattering 
of a kink on a 
trapped anti-kink (or vice-versa). If this process is possible, then it would 
provide a method to experimentally generate a breather from a kink and 
an anti kink. We plan to investigate this in the future.

\section{Acknowledgement}
The work reported in this paper was, in part, supported by a 
PPARC grant  (PPA/G/S/2003/00161).

\end{document}